\documentclass[prx, twocolumn, superscriptaddress, footinbib, longbibliography]{revtex4-1}

\usepackage{amsmath, amssymb, amsfonts}
\allowdisplaybreaks[4]

\usepackage{graphicx}
\graphicspath{ {./} {./figures/} {../figures/} {../../figures/} }
\usepackage[caption=false,subrefformat=parens,labelformat=parens]{subfig}
\usepackage{sidecap}

\usepackage{pdfpages}
\makeatletter
\AtBeginDocument{\let\LS@rot\@undefined}
\makeatother

\def\e{\mathrm{e}}
\def\d{\mathrm{d}}
\def\tr{\operatorname{tr}}

\begin{document}

\title{Geometry of Environment-to-Phenotype Mapping:\\
       Unifying Adaptation Strategies in Varying Environments}

\author{BingKan Xue}
\affiliation{The Simons Center for Systems Biology, Institute for Advanced Study, Princeton, NJ 08540}
\affiliation{Laboratory of Living Matter and Center for Studies in Physics and Biology, The Rockefeller University, New York, NY 10065}
\author{Pablo Sartori}
\affiliation{The Simons Center for Systems Biology, Institute for Advanced Study, Princeton, NJ 08540}
\affiliation{Laboratory of Living Matter and Center for Studies in Physics and Biology, The Rockefeller University, New York, NY 10065}
\author{Stanislas Leibler}
\affiliation{The Simons Center for Systems Biology, Institute for Advanced Study, Princeton, NJ 08540}
\affiliation{Laboratory of Living Matter and Center for Studies in Physics and Biology, The Rockefeller University, New York, NY 10065}

\date{\today}

\begin{abstract}
Biological organisms exhibit diverse strategies for adapting to varying environments. For example, a population of organisms may express the same phenotype in all environments (``unvarying strategy''), or follow environmental cues and express alternative phenotypes to match the environment (``tracking strategy''), or diversify into coexisting phenotypes to cope with environmental uncertainty (``bet-hedging strategy''). We introduce a general framework for studying how organisms respond to environmental variations, which models an adaptation strategy by an abstract mapping from environmental cues to phenotypic traits. Depending on the accuracy of environmental cues and the strength of natural selection, we find different adaptation strategies represented by mappings that maximize the longterm growth rate of a population. The previously studied strategies emerge as special cases of our model: the tracking strategy is favorable when environmental cues are accurate, whereas when cues are noisy, organisms can either use an unvarying strategy or, remarkably, use the uninformative cue as a source of randomness to bet-hedge. Our model of the environment-to-phenotype mapping is based on a network with hidden units; the performance of the strategies is shown to rely on having a high-dimensional internal representation, which can even be random.
\end{abstract}

\maketitle

\section{Introduction}

To study the properties of a physical system, a phenomenological approach is to characterize how it responds to external conditions. For instance, materials show particular patterns of deformation under external forces, which reveals their elastic properties. Biological organisms exhibit far more complex responses to environmental conditions. As the environment varies, organisms adapt by changing their phenotypes, including morphological and behavioral traits. Such phenotypic responses to the environment are modified through the process of evolution, which gives rise to different forms of adaptation. Several adaptation strategies, as described below, have been studied both experimentally and theoretically \cite{Kassen2002, Sexton2017, Slatkin1974, Simons2011, Grimbergen2015, DeWitt1998, Hendry2016, Mayer2017}. In this paper, we adopt the phenomenological approach to study biological adaptation by modeling general forms of phenotypic responses to environmental conditions. This approach enables us to reveal underlying connections between different adaptation strategies.

The simplest adaptation strategy is one in which organisms express the same phenotype in all environments. A population using this strategy has a narrow distribution of phenotypes that does not vary with the environment. In such an ``unvarying strategy'', the typical phenotype is often fit for most environmental conditions. For example, birds that feed on a variety of food sources (``generalists'') have a mid-sized beak, which is slender enough for catching insects and conical enough for cracking seeds \cite{Grant1986, Shoval2012}.

Another strategy is for organisms to follow environmental cues and express alternative phenotypes to match the environment. Provided that the cues are accurate, individual organisms of a population may all express the appropriate phenotype. The phenotype distribution would thus exhibit a narrow peak that tracks the environmental variation. Examples of this ``tracking strategy'' are seasonal changes of butterfly's wing patterns and mammal's coat colors, which are induced by weather conditions and provide suitable camouflage \cite{Moczek2011}.

A third strategy is such that individual organisms of the same population express different phenotypes, so that the phenotype distribution is broad or has multiple peaks. Such diversification is useful in stochastically changing environments, since there will always be some individuals in the population that have the right phenotype to survive. A classical example of this ``bet-hedging strategy'' is the seed bank: to cope with unpredictable inclement weather, some seeds quickly germinate after being dispersed while others remain in the soil for a prolonged period \cite{Cohen1966, Venable2007}. Bet-hedging can also be combined with cue-tracking, such that the distribution of phenotypes varies according to the environment. For example, the fraction of seeds that germinate can depend on environmental factors such as temperature, moisture, and the presence of other seeds \cite{Cohen1967, Gremer2016}.

We will show that the above strategies are special limits of a general solution for adaptation to varying environments. Depending on the accuracy of environmental cues and the strength of natural selection, particular strategies of adaptation emerge from a continuum of possible strategies. This unifying picture is obtained using a model of ``environment-to-phenotype mapping'', which allowed us to explore a wide range of phenotypic responses to environmental conditions. Essential to our model is a high-dimensional internal representation of the environment that allows organisms to develop diverse phenotypic responses. Our results suggest ways to experimentally evolve and identify different adaptation strategies.

\section{Model of environment-to-phenotype mapping}

The phenotypic responses of an organism to environmental conditions can be conceptualized as a mapping from the environment space to the phenotype space. A certain environmental stimulus that the organism experiences may induce a particular phenotype. Such an environment-to-phenotype mapping may represent, for example, how the development of organisms is affected by the environment (known as ``phenotypic plasticity''). A mapping that allows a population to survive better and reach greater abundance in the long term will generally be favored by natural selection. We will study the optimal form of the mapping that maximizes the population growth rate in varying environments.

Consider a population of organisms that reproduce asexually in discrete numbers of generations. The environment they live in may vary from generation to generation. An environmental condition will be described by an $n$-dimensional vector $\varepsilon$, whose components represent different environmental factors, such as temperature, light, and amount of food. We assume that the environment switches between several different conditions, labeled by $\varepsilon^\mu$ for $\mu = 1, \cdots, m$. Each individual organism receives an environmental cue, which is correlated with the environmental condition and can potentially be used to distinguish the actual environment. This environmental cue is denoted by a vector $\xi$, which is assumed to belong to the $n$-dimensional environment space. Note that, in the same environment $\varepsilon^\mu$, each organism may receive a different cue $\xi$.

\begin{figure*}
\centering
\includegraphics[]{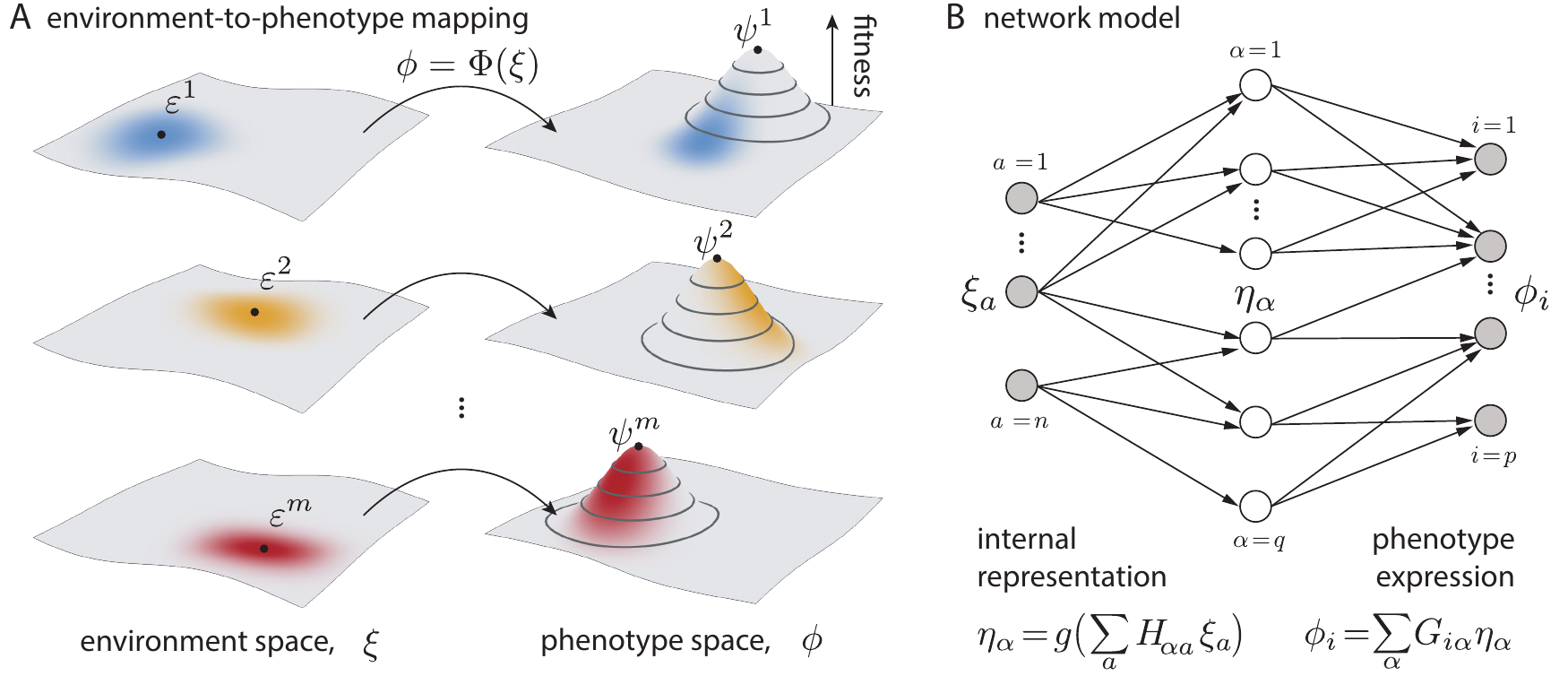}
\caption{\small Schematic illustration of our modeling framework: (A) Phenotypic responses described by a mapping from an $n$-dimensional environment space to a $p$-dimensional phenotype space. The environment can be in one of $m$ conditions, labeled by $\varepsilon^\mu$, each favoring a phenotype $\psi^\mu$. In a given environmental condition (distinguished by color), each individual organism receives a noisy cue $\xi$ (distribution represented by color shade in environment space), and expresses a phenotype according to the mapping $\Phi$ (distribution of phenotypes induced by the mapping is represented by color shade in phenotype space). The fitness of a phenotype depends on its distance to the favorable phenotype (illustrated by the fitness landscape in phenotype space). (B) A network model with one hidden layer. The input $\xi$ has $n$ components, $\xi_a$. The hidden layer has $q$ components, given by $\eta_\alpha = g(\sum_a H_{\alpha a} \xi_a)$, where $H_{\alpha a}$ is the representation matrix and $g$ is a sigmoid function. The output $\phi$ has $p$ components, determined by $\phi_i = \sum_\alpha G_{i\alpha} \eta_\alpha$, where $G_{i\alpha}$ is the expression matrix.}
\label{fig:schematics}
\end{figure*}

Similarly, the phenotype of an organism will be described by a $p$-dimensional vector $\phi$, whose components represent different characteristic traits, such as the shape of body parts or the speed of movement. The phenotype that an organism expresses may depend on the environmental cue $\xi$ that it receives. We will describe such dependence by a function, $\phi = \Phi(\xi)$, which represents a mapping from the $n$-dimensional environment space to the $p$-dimensional phenotype space, as illustrated in Fig.~\ref{fig:schematics}A. Different forms of the mapping will correspond to different adaptation strategies.

The fitness of an organism in a given environment $\varepsilon^\mu$ is measured by how many offspring it produces. This depends on its phenotype $\phi$, and will be described by a function $f(\phi;\varepsilon^\mu)$. Thus, in each generation, labeled by a number $t$, an individual organism that receives an environmental cue $\xi_t$ will express a phenotype $\phi_t = \Phi(\xi_t)$ and produce as many as $f(\phi_t;\varepsilon_t)$ offspring, where $\varepsilon_t$ is the environmental condition. Let $N_t$ be the population size in the $t$-th generation, then in the next generation it will be
\begin{equation} \label{eq:Nt}
N_{t+1} = N_t \, \sum_{\xi_t} P(\xi_t|\varepsilon_t) f(\Phi(\xi_t);\varepsilon_t) \,,
\end{equation}
where $P(\xi_t|\varepsilon_t)$ is the probability that a cue $\xi_t$ is received when the environment is $\varepsilon_t$. In the long term, the growth rate of the population is given by $\Lambda \equiv \frac{1}{T} \log \frac{N_T}{N_0}$ for $T \to \infty$. This long-term growth rate can be calculated as
\begin{equation} \label{eq:Lambda}
\Lambda = \sum_\mu p_\mu \log \sum_{\xi} P(\xi|\varepsilon^\mu) f(\Phi(\xi);\varepsilon^\mu) \,,
\end{equation}
where $p_\mu$ is the probability that each environmental condition $\varepsilon^\mu$ occurs. We will use $\Lambda$ as the measure of evolutionary success for a population. The optimal phenotypic response will be determined by the function $\Phi$ that maximizes the value of $\Lambda$.

For simplicity, we assume that the environmental cue $\xi$ is randomly distributed around the actual environment $\varepsilon^\mu$ according to a Gaussian distribution, $P(\xi|\varepsilon^\mu) = \frac{1}{(2\pi \sigma^2)^{n/2}} \exp\{ -\frac{(\xi - \varepsilon^\mu)^2}{2 \sigma^2} \}$, where $\sigma$ represents the noisiness of the environmental cue. The fitness is also assumed to be a Gaussian function, $f(\phi;\varepsilon^\mu) = F_\mu \exp\{ -\frac{\gamma^2 (\phi - \psi^\mu)^2}{2} \}$, where $F_\mu$ is a constant representing the maximum number of offspring in the environment $\varepsilon^\mu$, and $\psi^\mu$ is the most favorable phenotype in that environment. The parameter $\gamma$ represents the strength of natural selection, which is assumed to be the same for all environments (see Suppl. Fig.~S3 for a different case). Note that $\sigma$ and $1/\gamma$ serve as characteristic scales for the environment and the phenotype space, respectively. Under those assumptions, the long-term growth rate $\Lambda$ is evaluated numerically according to Eq.~[\ref{eq:Lambda}], as described in Appendix.

We are interested in the ideal function $\Phi^*$ that maximizes $\Lambda$, which satisfies the variational equation $\delta\Lambda / \delta\Phi(\xi) = 0$. Unfortunately, this equation cannot be solved explicitly in general (but see Appendix for special cases). To proceed further, we need to specify the function $\Phi$ in a parametric form, so that we can optimize over the parameters numerically. The form of the function should be sufficiently general in order to allow all possible types of phenotypic responses. In the following, we introduce a particular form of the function that is biologically motivated as well as computationally convenient.

Our model of the function $\Phi$ takes the form of a feed-forward network with a hidden layer. The input layer has $n$ nodes, corresponding to the $n$ components of the environmental cue $\xi$; the output layer has $p$ nodes, corresponding to the $p$ components of the phenotype $\phi$; the hidden layer is chosen to have $q$ nodes, a potentially large number compared to $n$ and $p$, as illustrated in Fig.~\ref{fig:schematics}B. These hidden nodes can be thought to form an internal representation of the external environment; their values are determined by the input vector $\xi$ through a ``representation matrix'' $H$ and a nonlinear transformation $g$, such as a $\tanh$ function. The output vector $\phi$ depends on the internal variables through an ``expression matrix'' $G$. All together, the function $\Phi$ takes the form
\begin{equation} \label{eq:network}
\phi_i = \Phi_i(\xi) = \sum_{\alpha}^{} G_{i\alpha} \, g \Big( \sum_{a}^{} H_{\alpha a} \xi_a \Big) \,.
\end{equation}
(Each matrix has an additional column that represents a constant (``bias'') term; e.g., $\sum_{a} H_{\alpha a} \xi_a \equiv \sum_{a=1}^{n} H_{\alpha a} \xi_a + H_{\alpha 0}$, where $H_{\alpha 0}$ is the constant term that is optimized as part of the matrix.)
With sufficiently many internal variables, such a multi-layered feed-forward network (known as a ``perceptron'' \cite{Hertz1991}) can approximate any smooth function and hence capture all possible phenotypic responses.

The structure of this model is inspired by many biological systems. The hidden nodes of the network may represent internal variables of the organism. For example, a plant's phenotypic responses to environmental conditions can be described by a growth-regulatory network, where a large group of molecules, such as growth factors and gene promoters, act as hidden nodes of the network \cite{Scheres2017}. The formation of a high-dimensional internal representation, which allows organisms to better perceive the environment and produce more refined phenotypic responses, has also been suggested. Cellular signaling networks, for example, involve many proteins that often have multiple modification sites, interacting with each other and giving rise to a large number of possible states \cite{Hlavacek2003}. Similarly, biological neural networks, such as the olfactory systems of insects and mammals, have multiple layers of neurons for processing sensory information; some intermediate layers of neurons may play the role of expanding the dimensionality of input signals to facilitate later stages of cognition \cite{Babadi2014, Krishnamurthy2017}.

In our network model, the environment-to-phenotype mapping is specified by the representation matrix $H$ and the expression matrix $G$. These matrices may represent information that are encoded in the organism's genotype, which undergoes evolution. For simplicity, we consider the case where individuals of the population share the same matrices, and we look for the optimal values of $H$ and $G$ that maximize the long-term population growth rate $\Lambda$.

\section{Emergence of different adaptation strategies}

\begin{SCfigure*}
\begin{minipage}[c]{4.1in}
\includegraphics[width=2in]{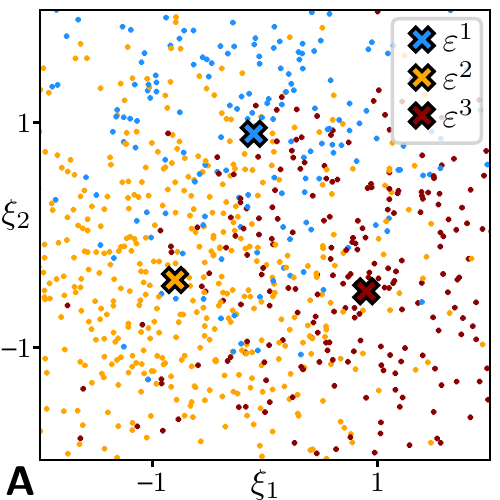}
~
\includegraphics[width=2in]{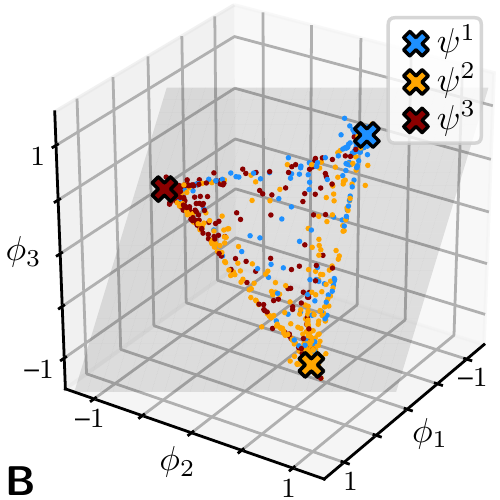}
\end{minipage}
\caption{\small Example of an adaptation strategy produced by an optimized network: (A) Distribution of environmental cues $\xi$ represented by points in the environment space (color represents the actual environmental condition $\varepsilon^\mu$). (B) Distribution of phenotypes produced by the optimized network, represented by points in the phenotype space. All points fall on a plane (gray transparent) spanned by the archetypes $\psi^\mu$. For these figures we used parameter values $\sigma = 1$ for environmental noise and $\gamma = 1$ for selection strength, which represent characteristic scales that are of the same order as the distance between two environments $\varepsilon^\mu$ and between two archetypes $\psi^\mu$, respectively.}
\label{fig:space}
\end{SCfigure*}

The adaptation strategy resulting from the optimized network will depend on the level of environmental noise $\sigma$ and the strength of natural selection $\gamma$. We explore the range of adaptation strategies in the $(\sigma, \gamma)$ parameter space using numerical examples. Consider a 2-dimensional environment space ($n=2$), a 3-dimensional phenotype space ($p=3$), and a 20-dimensional internal space ($q=20$). The environment switches between three conditions ($m = 3$), with arbitrarily chosen positions in the environment space (marked in Fig.~\ref{fig:space}A) and probabilities of occurrence ($p_\mu = 0.2$, $0.5$, and $0.3$, respectively). For each environmental condition $\varepsilon^\mu$, we assign a most favorable phenotype $\psi^\mu$, called ``archetype'' hereafter, in the phenotype space (Fig.~\ref{fig:space}B). In a given environment $\varepsilon^\mu$, organisms receive a distribution of cues, as illustrated in Fig.~\ref{fig:space}A. The mapping given by the optimized network generates a distribution of phenotypes, as illustrated in Fig.~\ref{fig:space}B. The shape of the phenotype distribution, and how it changes under different environmental conditions, characterizes the corresponding adaptation strategy.

A prominent feature of the emerged geometric structure, shown in Fig.~\ref{fig:space}B, is that all phenotypes lie on a flat plane spanned by the archetypes, $\{ \psi^\mu \}$. This structure can be explained by a ``Pareto efficiency'' argument as follows. Since the fitness of a phenotype depends on its distance to the archetypes, a phenotype located off the plane will always be less fit than its perpendicular projection onto the plane. Therefore, in the optimal phenotype distribution, all phenotypes should fall on the plane. In general, if there are $m$ archetypes, the optimal phenotype distribution will be contained in a $(m-1)$ dimensional subspace spanned by those archetypes. If $m$ is small compared to the dimensionality of the original phenotype space, $p$, then the dimensionality of phenotypes will be significantly reduced.

Such dimensional reduction, as well as the Pareto efficiency argument, is similar to that found in the model of Ref.~\cite{Shoval2012}. In that model, the archetypes represent different biological tasks that every individual organism must perform during its lifetime, with varied degrees of importance to its overall fitness. To compare with our model, we can associate the tasks with environmental conditions that individuals may encounter and need adapt to, with varied probabilities of occurrence. From this perspective, the model of Ref.~\cite{Shoval2012} corresponds to the situation where the phenotype does not depend on the present environment (i.e., no phenotypic plasticity), and the phenotype distribution of a population is simply localized at a given point in the phenotype space. This form of phenotypic response and the resulting phenotype distribution are characteristic of the unvarying strategy, which will be discussed later. In contrast, by allowing the phenotype to depend on environmental cues through the environment-to-phenotype mapping, our model encompasses a wider range of adaptation strategies, as we shall see below.

\begin{figure*}
\centering
\includegraphics[]{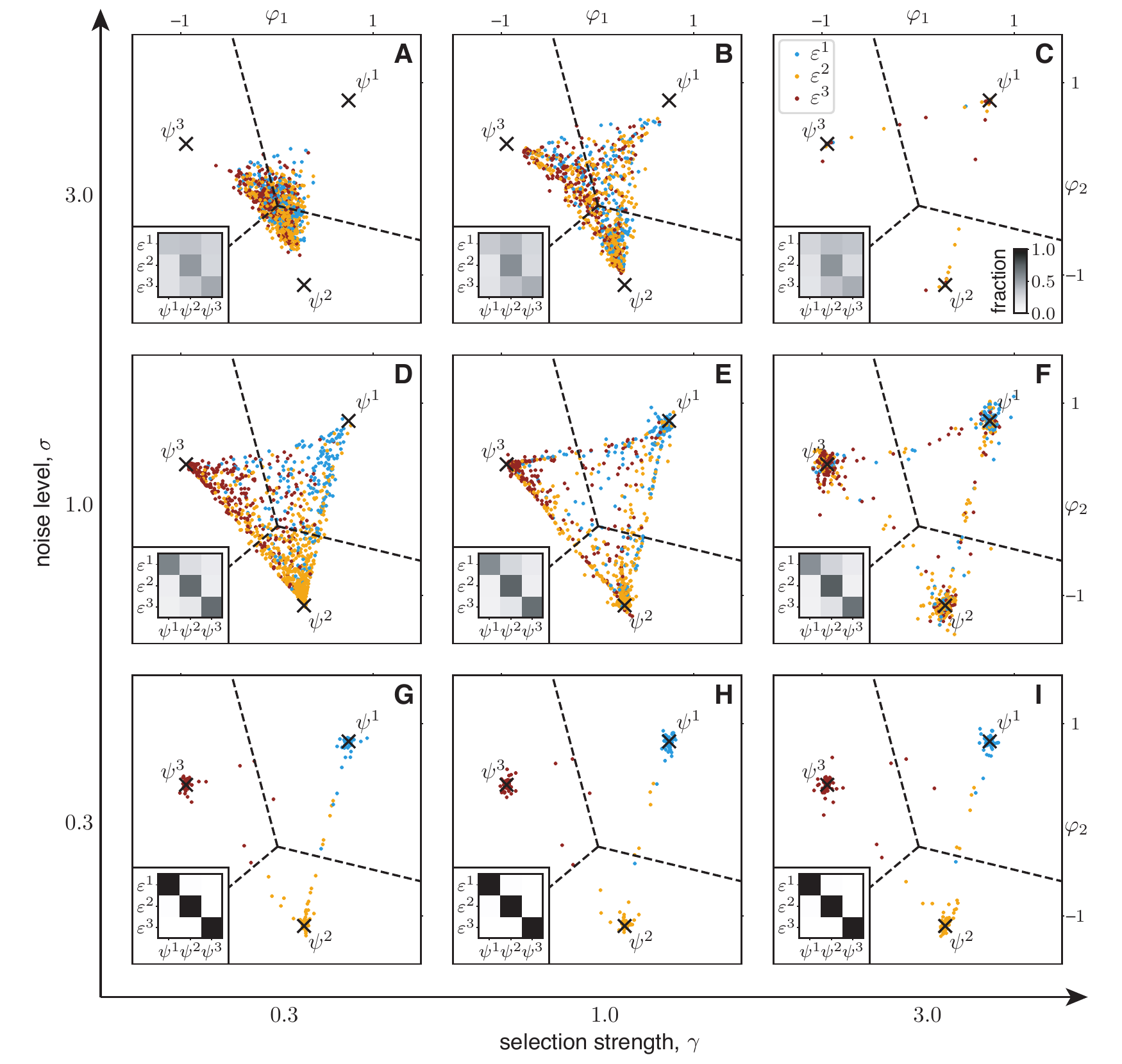}
\caption{\small Phenotype distributions produced by networks optimized for different values of the noise level $\sigma$ and the selection strength $\gamma$. Color dots represent phenotypes plotted in the archetype plane, with $\varphi_1, \varphi_2$ as new coordinates. Dashed lines divide the plane into regions that are close to each $\psi^\mu$; the intersection point corresponds to the average phenotype, $\bar{\psi} = \sum_\mu p_\mu \psi^\mu$. Insets show the fraction of phenotypes inside each region under different environmental conditions.}
\label{fig:phe2d}
\end{figure*}

\begin{figure*}
\centering
\includegraphics[]{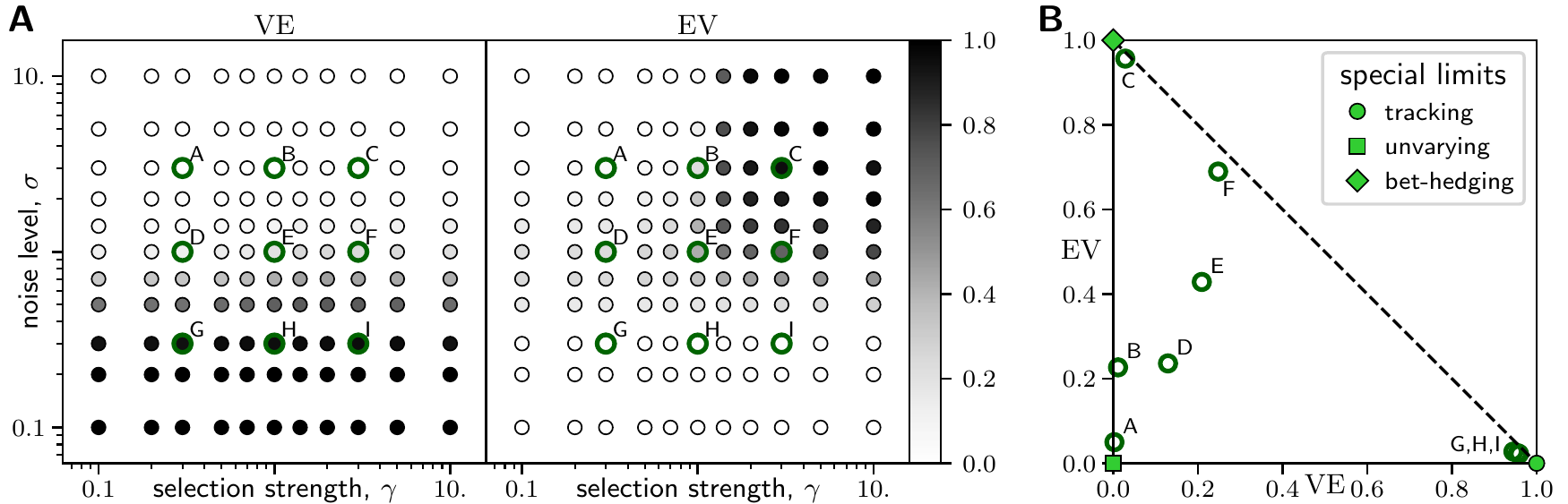}
\caption{\small Characterization of adaptation strategies using quantities $\mathrm{VE}$ and $\mathrm{EV}$: (A) Plot of the parameter space showing how $\mathrm{VE}$ and $\mathrm{EV}$ vary with the noise level $\sigma$ and the selection strength $\gamma$. Circles represent data points from numerical calculations, with values of $\mathrm{VE}$ and $\mathrm{EV}$ illustrated by gray-scale; thick colored circles correspond to examples shown in Fig.~\ref{fig:phe2d}. (B) Plot of $\mathrm{VE}$-$\mathrm{EV}$ space showing examples from Fig.~\ref{fig:phe2d}, with the same colors as in (A). Dashed line represents the bound $\mathrm{VE} + \mathrm{EV} \leq 1$. Corners of the $\mathrm{VE}$-$\mathrm{EV}$ space represent special limits that correspond to the tracking, unvarying, and bet-hedging strategies.}
\label{fig:order}
\end{figure*}

\subsection{Examples of strategies}

In the following, we examine the distribution of phenotypes for different parameters $\sigma$ and $\gamma$, represented by the density of points in the archetype plane, as shown in Fig.~\ref{fig:phe2d} (also see Suppl. Fig.~S1 for clarity). In many cases, the density is high near the archetypes. We divide the plane into regions surrounding each $\psi^\mu$, marked by boundary lines in Fig.~\ref{fig:phe2d}; the fraction of phenotypes lying inside each region is shown in the insets. By comparing those fractions as well as the shape of the phenotype distribution between different environmental conditions, we identify a wide range of adaptation strategies.

\textbf{Tracking strategy under low noise.}
Examples of low environmental noise are shown in Figs.~\ref{fig:phe2d}G--I. In these cases, the width of the noise distribution is much smaller than the typical distance between two environmental conditions (chosen to be $\simeq 1$), i.e., $\sigma \ll 1$. Therefore, the environmental cue is very accurate about the present environmental condition. As a result, in each environment $\varepsilon^\mu$, the phenotype distribution is highly concentrated near the corresponding archetype $\psi^\mu$ --- the surrounding region contains almost 100\% of the phenotypes, so the inset plot looks diagonal. This means that the organisms can express the most favorable phenotype that tracks the varying environmental condition. The picture hardly changes as the selection strength $\gamma$ is varied (compare Figs.~\ref{fig:phe2d}G--I). It is understandable since, without a significant cost for sensing, organisms should always utilize environmental cues when those are reliable.

\textbf{Unvarying strategy under high noise and weak selection.}
The opposite case where environmental noise level is high ($\sigma \gg 1$) is shown in Figs.~\ref{fig:phe2d}A--C. In these examples, the environmental cue has a broad distribution and is largely uninformative about the actual environment. Therefore, we expect the optimal phenotype distributions to look similar in all environments. This is verified by Figs.~\ref{fig:phe2d}A--C, where the insets show that there is a significant fraction of phenotypes in each region and the fractions vary slightly between different environments (see also Suppl. Fig.~S1). However, depending on the selection strength $\gamma$, the phenotype distribution has very different characters. Fig.~\ref{fig:phe2d}A shows the case of weak selection, where the characteristic scale $1/\gamma$ is much larger than the typical distance between two phenotypes (chosen to be $\simeq 1$), i.e., $\gamma \ll 1$. In this case, the phenotypes are centered near the average phenotype, $\bar{\psi} = \sum_\mu p_\mu \psi^\mu$, regardless of the environmental condition. It means that the organisms have evolved to ignore the cue when it is noisy and exhibit a constant phenotype. The optimal constant phenotype strikes a balance between all the archetypes, similar to the result of Ref.~\cite{Shoval2012}.

\textbf{Bet-hedging strategy under high noise and strong selection.}
When the cue is noisy and the selection is strong ($\sigma, \gamma \gg 1$), however, the unvarying strategy fails because the average phenotype $\bar{\psi}$ suffers from low fitness values in all environments. In this case, surprisingly, the organisms do not ignore the uninformative environmental cue, but use it in a completely different way --- each organism expresses one of the archetypes according to the cue, so that the population diversifies into multiple subpopulations due to the randomness of the cue. As shown in Fig.~\ref{fig:phe2d}C, the phenotype distribution is sharply peaked around every archetype $\psi^\mu$, and the size of each peak changes little with the environmental condition. This bet-hedging strategy guarantees that, in any environment $\varepsilon^\mu$, a subpopulation expressing the corresponding archetype $\psi^\mu$ will have a high fitness value. The relative size of each subpopulation depends on the probability $p_\mu$ that each environment occurs. In the limit of extremely strong selection ($\gamma \to \infty$), we expect to recover the result of previous bet-hedging models (e.g., \cite{Rivoire2011}), in which the probability of expressing the archetype $\psi^\mu$ matches the probability of encountering the environment $\varepsilon^\mu$. This is indeed the case, as the fraction of phenotypes near each $\psi^\mu$ agrees well with the environment probability $p_\mu$ (see also Suppl. Fig.~S1C).

\textbf{Intermediate strategies.} Besides the above extreme cases that correspond to well categorized adaptation strategies, intermediate cases are also found. A combination of bet-hedging and tracking strategies is seen in the case of a medium noise level ($\sigma \simeq 1$) and strong selection ($\gamma \gg 1$). As shown in Fig.~\ref{fig:phe2d}F, the phenotype distribution is peaked around the archetypes, but the relative sizes of the peaks are biased towards the one that matches the actual environment (see also Suppl. Fig.~S1F). This case may represent the situation of bet-hedging with partial environmental information, in which the population uses an imperfect cue to moderately adjust its phenotype distribution \cite{Rivoire2011, Donaldson-Matasci2013}. Similarly, we can see intermediate cases between bet-hedging and unvarying strategies (high noise $\sigma \gg 1$ and medium selection $\gamma \simeq 1$, Fig.~\ref{fig:phe2d}B), as well as between unvarying and tracking strategies (medium noise $\sigma \simeq 1$ and weak selection $\gamma \ll 1$, Fig.~\ref{fig:phe2d}D).

The transition of adaptation strategies with the parameters $\sigma$ and $\gamma$, illustrated by the examples in Fig.~\ref{fig:phe2d}, can also be understood analytically using approximate solutions of the ideal function $\Phi^*$ for extreme parameter values (see Appendix). Those approximate solutions do not rely on the parametric form of the function, Eq.~(\ref{eq:network}), showing that our results are more general than the numerical examples. Generally, the accuracy of environmental cues, measured by the noise level $\sigma$, determines the bias of the phenotype distribution towards the archetype in a given environmental condition. The selection strength $\gamma$, on the other hand, modifies the shape of the phenotype distribution, which tends to be more clustered near the archetypes when the selection is strong, and more scattered into the interior space between the archetypes when the selection is weak.

\subsection{Quantification of strategies}

The shape of the phenotype distributions illustrated above can be characterized quantitatively. Two main properties of the phenotype distributions are how much they vary with the environment and how concentrated they are near the archetypes. To describe these properties, we introduce two characteristic quantities and examine how they vary with the environmental noise $\sigma$ and the selection strength $\gamma$.

Specifically, in each environment $\varepsilon^\mu$, the phenotype distribution can be denoted by a conditional probability distribution $\pi(\phi|\varepsilon^\mu)$, as defined in Eq.~(\ref{eq:pi|mu}). Given the environment probabilities $p_\mu$, the overall distribution of the phenotype is $\pi(\phi) = \sum_\mu p_\mu \pi(\phi|\varepsilon^\mu)$. The total variance of the phenotype can be decomposed as $\mathbb{V}[\phi] = \mathbb{V}[\mathbb{E}[\phi|\varepsilon^\mu]] + \mathbb{E}[\mathbb{V}[\phi|\varepsilon^\mu]]$. In the first term, $\mathbb{E}[\phi|\varepsilon^\mu]$ is the conditional expectation of the phenotype for a given environment $\varepsilon^\mu$, and $\mathbb{V}[\mathbb{E}[\phi|\varepsilon^\mu]]$ is the variance of the conditional expectation with respect to the environment probabilities $p_\mu$; and similarly for the second term. We can use these two terms to characterize different adaptation strategies. Essentially, the first term characterizes how much the phenotype varies with the environment, whereas the second term characterizes how much the phenotype varies in a given environment. For clarity, we take the trace of the variance matrices and normalize the terms by the variance of the archetypes, $\mathbb{V}[\psi]$ (according to the Pareto efficiency argument, the optimal phenotype distributions are contained in between the archetypes, hence $\mathbb{V}[\phi] \leq \mathbb{V}[\psi]$). Thus, our characteristic quantities are
\begin{equation}
\mathrm{VE} \equiv \frac{\tr \big(\mathbb{V} \big[ \mathbb{E}[\phi|\varepsilon^\mu] \big] \big)}{\tr \big( \mathbb{V}[\psi] \big)} \,, \quad
\mathrm{EV} \equiv \frac{\tr \big(\mathbb{E} \big[ \mathbb{V}[\phi|\varepsilon^\mu] \big] \big)}{\tr \big( \mathbb{V}[\psi] \big)} \,.
\end{equation}
Fig.~\ref{fig:order}A shows how the values of these quantities change according to the parameters $\sigma$ and $\gamma$.

To see how these quantities help characterize different adaptation strategies, consider the three strategies described above. For the tracking strategy, the phenotypes are concentrated near the corresponding archetype in each environment, hence $\mathbb{E}[\phi|\varepsilon^\mu] \approx \psi^\mu$ and $\mathbb{V}[\phi|\varepsilon^\mu] \approx 0$; therefore, $\mathrm{VE} \approx 1$ and $\mathrm{EV} \approx 0$. Similarly, for the unvarying strategy, the phenotypes are always concentrated near the center of the archetypes, which means $\mathbb{E}[\phi|\varepsilon^\mu] \approx \bar{\psi}$ and $\mathbb{V}[\phi|\varepsilon^\mu] \approx 0$; therefore, $\mathrm{VE} \approx 0$ and $\mathrm{EV} \approx 0$. Finally, for the bet-hedging strategy, the phenotype distributions are largely independent of the environment, and are concentrated near the archetypes in proportion to the environment probabilities $p_\mu$; this leads to $\mathrm{VE} \approx 0$ and $\mathrm{EV} \approx 1$. Therefore, those three strategies can be clearly distinguished by different limits of the characteristic quantities, as shown in Fig.~\ref{fig:order}B.

\section{Dimensionality of internal representation}

So far we have fixed the dimensionality of the network's hidden layer at a relatively large number, $q = 20$, as compared to that of the environment space, $n = 2$. The motivation was to create an adequate expansion of dimensionality from the input layer to the hidden layer, $q/n = 10$, so that the network can be used to approximate well the ideal function $\Phi^*$ in all cases. The approximation is verified in the limit $\gamma \to 0$, where explicit solutions can be found (see Appendix); the numerical solutions we obtained are very close to the ideal function $\Phi^*$ (Suppl. Fig.~S2B--C).

\begin{figure}
\centering
\includegraphics[]{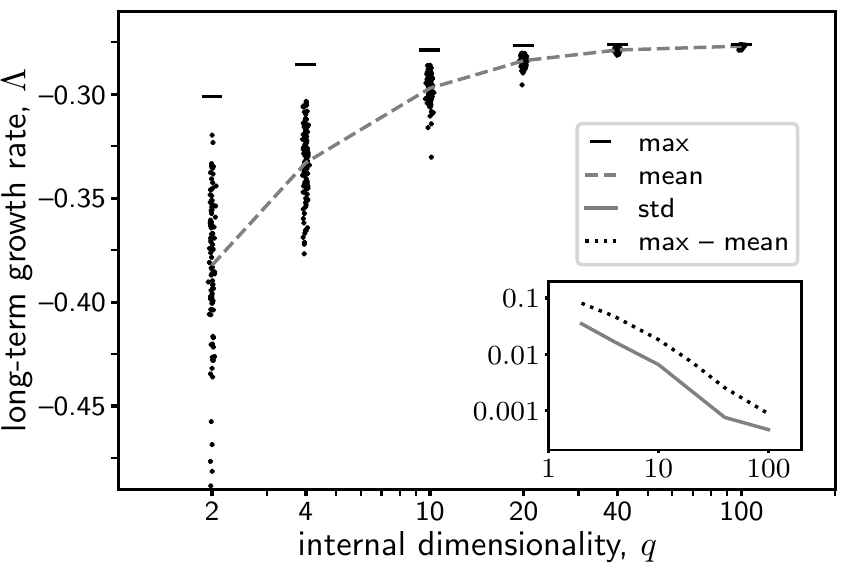}
\caption{\small Long-term population growth rate $\Lambda$ versus the dimensionality $q$ of the intermediate layer of the network. Each point represents a network with a random, fixed representation matrix $H$ (entries drawn from $\mathcal{N}(0,1)$ independently) and an optimized expression matrix $G$. (To aid visualization of the density of points, a small random horizontal displacement is added.) Dashed and solid (inset) lines show the mean and standard deviation of the values of $\Lambda$. Horizontal bars mark the maximum values of $\Lambda$ when the matrix $H$ is also optimized for each $q$; dotted line (inset) shows the difference between the maximum and the mean values. For this example the parameter values $\sigma = 1$ and $\gamma = 1$ are used.}
\label{fig:fitness}
\end{figure}

Let us now explore how the results change if we vary the dimensionality $q$. Fig.~\ref{fig:fitness} shows how the maximum value of $\Lambda$ increases with $q$. For a small $q$, the network model becomes very restrictive because it does not have many parameters that can be tuned. In that case, the phenotype distribution that results from optimizing the network will be deformed from that for the ideal function $\Phi^*$ (see Suppl. Fig.~S2A). In particular, in the limit $q \to 0$, the intermediate layer of the network vanishes, so the output becomes disconnected from the input. This means that the phenotype can no longer depend on the environmental cue, hence the organism is forced to express the same phenotype in all environments. In other words, the organism can only use the unvarying strategy, even though it is not favorable in many situations. On the other hand, a large $q$ enables organisms to form various types of adaptation strategies, as we have seen for $q = 20$. The price, however, is having to tune a lot of parameters. This could mean a much longer time for a population to adapt to a varying environment.

In our numerical computation, we found that it is much slower to optimize over the representation matrix $H$ than the expression matrix $G$, because the latter is directly connected to the output phenotype being selected but the former is not. This suggests that it is harder for an organism to adjust the way it creates an internal representation of the environment than to adjust the mechanism that produces the phenotype directly. It is therefore interesting to ask if one can keep the representation matrix $H$ fixed while optimizing over the expression matrix $G$ alone.

To address this point, we consider the case where the representation matrix is chosen randomly. For a given dimensionality $q$, let each entry of $H$ be drawn independently from a standard normal distribution $\mathcal{N}(0,1)$. For each of such random, fixed matrix $H$, the network is optimized over $G$ to maximize the long-term population growth rate $\Lambda$. The results are shown in Fig.~\ref{fig:fitness}. We find that, for a relatively small $q$ (such as $q = 4$), the values of $\Lambda$ are low and widely spread; however, for a very large $q$ (such as $q = 100$), the values of $\Lambda$ are not only high but also narrowly distributed. Moreover, the distribution of $\Lambda$ values moves closer to the maximum value as the dimensionality $q$ increases. Hence, with a sufficiently high dimensionality, a random representation can be almost as good as the optimal one. This suggests that having a high-dimensional, sufficiently complex, internal representation of the environment would allow organisms to flexibly and quickly adapt to many situations. Of course, maintaining a large number of internal variables may incur additional costs.

The idea that a high-dimensional and potentially random representation of the input can encode complicated output patterns is related to the kernel method and reservoir computing in machine learning \cite{Lukosevicius2009}. In general, more complex patterns require higher dimensionality of the internal representation (see \cite{Hertz1991, Mohri2012} for discussion on the limitation of such methods). Similar ideas have been explored in biological contexts \cite{Babadi2014, Krishnamurthy2017}.

\section{Discussion}

We have presented a general model of organisms' phenotypic responses to varying environments; the optimal responses show patterns of adaptation observed in nature. The form of such adaptation strategies depends on the noisiness of environmental cues and the selectivity of environmental conditions. In special limits of the parameter values, we have recovered three well-known strategies --- unvarying, bet-hedging, and tracking. The capacity of forming these and other adaptation strategies depends on the richness of the organisms' internal representation of the environment, characterized in our model by the number of internal variables.

\subsection{Separation of timescales}

Our model implicitly assumes the separation of characteristic timescales of phenotypic responses, environmental changes, and evolution. In particular, by considering time in discrete numbers of generations, we do not model explicitly the dynamics of phenotypic development and environmental changes within a generation. This simplification is easily understood in cases where the timescale of environmental changes is much longer than that of the developmental process. In other cases, where the environment and the phenotype vary significantly within the lifetime, the vectors $\varepsilon$ and $\phi$ can in principle represent time courses of the environment and the phenotype, respectively, such as growth conditions and behavioral traits during the lifetime of an organism. This would naturally make those vectors high dimensional and the mapping more complicated, which may inspire additional consideration on modeling the dynamics of phenotypic responses.

We have also assumed that the timescale of environmental changes is much shorter than that of evolutionary changes. This allowed us to consider the effect of evolution in varying environments by optimizing the environment-to-phenotype mapping with respect to the environmental statistics, without explicitly treating the dynamics of the evolutionary process. It should be noted that, when the timescale of environmental changes is comparable to that of evolutionary changes (such as the time for genetic mutations to arise and spread in a population), different modes of evolutionary dynamics may occur. Such situations have been theoretically studied in models of population genetics. For example, during a prolonged period of constant environment, organisms may lose the plasticity to express alternative phenotypes due to the accumulation of mutations affecting unused phenotypes \cite{Gerland2009, Masel2007}. Similarly, bet-hedging can be selected against in such a situation \cite{King2007}, and the population could go extinct before profiting from environmental changes.

When the environment is correlated over multiple generations, it is possible to reduce uncertainty in estimating the environment by tracking the history of environmental cues. This can be done by having organisms pass down information about their environment to their offspring, e.g., through epigenetic inheritance. Our current model does not include such possibility, since the phenotype of an organism depends only on the environmental cue it receives, and not on its parent's cue or phenotype. To incorporate transgenerational effects, one could, for example, let the state of the network in one generation depend on that in the previous generation, thus making the network recurrent across generations. Such generalization would allow the organisms to utilize temporal structures in the environmental variation.

\subsection{Relation to experiments}

The geometry of phenotypic responses associated with different adaptation strategies can be looked for in experimental studies. Such studies should involve measuring the phenotype distribution in a wide range of controlled environmental conditions. Each strategy may be recognized by a particular shape of the phenotype distribution. For instance, an unvarying strategy is characterized by a phenotype distribution with a single peak that is stable under environmental variations. A pure bet-hedging strategy is associated with a multi-modal phenotype distribution that does not depend on the environment. A tracking strategy, on the other hand, features a phenotype distribution with a single peak that changes position according to the environmental condition.

Our model predicts that specific adaptation strategies emerge under different levels of environmental noise and selection pressure. These predictions can be tested by experimental evolution. Indeed, several experiments have demonstrated that particular forms of adaptation can be evolved. For example, phenotypic plasticity, crucial for the tracking strategy in which organisms express distinctive phenotypes under varied environmental conditions, has been observed in larval development under temperature treatments \cite{Suzuki2006}. The evolution of bet-hedging strategies has been shown in bacteria subject to repeated selection in contrasting growth conditions \cite{Beaumont2009}. The random choice of phenotypes in a bet-hedging strategy may come from stochasticity in biochemical processes inside the organism. Alternatively, our model suggests that, when environmental cues are noisy and selection is strong, organisms can evolve to bet-hedge using the cue as a source of randomness. Remarkably, a recent experiment in yeast showed that, indeed, bet-hedging can be generated by plastic responses to an uninformative cue \cite{Maxwell2017}. Ultimately, a full test of our model requires varying the noise level of environmental cues and selection strength of environmental conditions, and showing that different patterns of adaptation emerge from evolution. Such experiments would require quantitative and systematic measurements of the relation between organisms' phenotype and their environment.

\section*{Conclusion}

We have introduced here the environment-to-phenotype mapping as an effective approach for studying the response of organisms to environmental conditions. This approach allowed us to explore a wide range of possible responses beyond the details of underlying molecular mechanisms. Compared to the commonly studied genotype-to-phenotype mapping, which describes how genetic variation affects phenotypes and emphasizes a mechanistic perspective \cite{Alberch1991, Pigliucci2010, Ahnert2017}, the environment-to-phenotype mapping provides a phenomenological perspective by describing organisms as a set of input-output relations that can be measured in experiments. This description is potentially useful for studying evolution, since the same form of phenotypic responses may be naturally selected even if it is implemented by different molecular mechanisms. For instance, many bacteria can stochastically switch from a normal growth state to a dormant persister state, which prevents cell death from unforeseeable antibiotic attack \cite{Balaban2004}. Different molecular mechanisms have been found to underly such bacterial persistence \cite{Cohen2013}. Nevertheless, the growth benefit of this particular adaptation strategy can be understood without using those mechanistic details \cite{Kussell2005a}. Such methods have recently been applied to other types of adaptation strategies \cite{Rivoire2011, Xue2018}.

We have used a network model as a simple example of possible forms of the environment-to-phenotype mapping. In our model the connections of the network store information about the environmental conditions and their statistics, as well as about the favorable phenotypes. Besides varying the dimensionality of the internal representation or the number of intermediate layers \cite{Friedlander2015}, a possible further generalization of our model would be to consider a recurrent network with evolvable internal dynamics \cite{Kaneko2012a}. Such a network could allow organisms to store information about their past phenotypes and encode temporal structures of the environmental history. The environment for the organisms can also include ecological interactions with individuals of the same population or other species. Such generalizations could lead to potentially more complex adaptation strategies.

\section*{Acknowledgments}

We thank Michael R. Mitchell, David A. Huse, Kunihiko Kaneko, and Lai-Sang Young for helpful discussions. This research has been partly supported by grants from the Simons Foundation to S.L. through the Rockefeller University (Grant No. 345430) and the Institute for Advanced Study (Grant No. 345801). B.X. is funded by the Eric and Wendy Schmidt Membership in Biology at the Institute for Advanced Study.

\appendix

\renewcommand{\theequation}{A\arabic{equation}}
\setcounter{equation}{0}
\renewcommand{\thefigure}{A\arabic{figure}}
\setcounter{figure}{0}
\renewcommand{\paragraph}[1]{
\medskip \noindent
\textbf{\normalsize #1}
\medskip
}

\section*{Appendix}

\subsection{Numerical methods} \label{app:numerical}

Our goal is to maximize the long-term growth rate $\Lambda$ with respect to the phenotypic response function $\Phi$. The function $\Phi$ is parametrized by the matrices $H$ and $G$, as in Eq.~[\ref{eq:network}] of the main text. The value of $\Lambda$, according to Eq.~[\ref{eq:Lambda}], is given by
\begin{align} \label{eq:Sint}
&\Lambda = \sum_\mu p_\mu \log F_\mu + \sum_\mu p_\mu \log \Big\langle \e^{-\frac{\gamma^2}{2} (\Phi(\xi) - \psi^\mu)^2} \Big\rangle_{\xi \sim \mathcal{N}(\varepsilon^\mu, \sigma)}
\end{align}
where $\langle\cdot\rangle$ represents the expectation value with respect to the Gaussian random variable $\xi$. The first term does not depend on the parameters of $\Phi$ and will be ignored. The optimization is done numerically by iterating over two steps: calculating the expectations in Eq.~[\ref{eq:Sint}] given the current values of $H$ and $G$, and updating these matrices to improve the value of $\Lambda$.

For the first step, we calculated the expectation values by numerically integrating over the Gaussian distributions. We used the python package ``scipy.integrate'', which calls the Fortran library QUADPACK. An alternative approach to numerical integration is to generate a random sample of $\xi$ from the Gaussian distribution and use it to estimate the expectation values. This approach represents a finite sampling of the environmental cues, which allows for the analysis of the effect of finite population sizes and the stability of the optimal solutions. We tried both approaches and did not find significant differences in performance.

For the second step, we searched parameters using the python package ``scipy.optimize'' with the Broyden-Fletcher-Goldfarb-Shanno (BFGS) algorithm. This step involves calculating the gradient of the function $\Lambda$ over the matrices $H$ and $G$, then using the gradient to update their values. One could update the matrices simultaneously, or optimize one while holding the other fixed and then iterate. It turns out that optimizing the matrix $G$ alone is efficient, because $G$ is directly connected to the output without having a nonlinear transformation. Using this observation, we chose to optimize $G$ at every step of updating $H$. In this case, the gradient of $\Lambda(G^*(H),H)$ over $H$ can be simply calculated as $\frac{\partial\Lambda}{\partial H}\big|_{G^*}$ because $\frac{\partial\Lambda}{\partial G}\big|_{G^*} = 0$.

For the examples shown in Fig.~\ref{fig:phe2d}, the coordinates of the environments and the archetypes are $\varepsilon^1 = [-0.1, 0.9]$, $\varepsilon^2 = [-0.8, -0.4]$, $\varepsilon^3 = [0.9, -0.5]$, $\psi^1 = [-0.6, 0.5, 0.8]$, $\psi^2 = [0.4, 0.6, -0.9]$, $\psi^3 = [0.5, -0.8, 0.4]$; the environment probabilities are $[p_1, p_2, p_3] = [0.2, 0.5, 0.3]$. The same values are used for Figs.~\ref{fig:order} and \ref{fig:fitness}. In Fig.~\ref{fig:order}, for each pair of parameter values $\sigma$ and $\gamma$, we ran 8 replicate optimizations starting from random initial values (every entry of $H$ and $G$ drawn i.i.d. from $\mathcal{N}(0,1)$); the order parameters are averaged over these replicates. In Fig.~\ref{fig:fitness}, for each dimensionality $q$, we ran 100 examples, each having a fixed $H$ with random entries.

\subsection{Analytic limits} \label{app:analytic}

Nonparametrically, the ideal response function $\Phi^*$ that maximizes Eq.~(\ref{eq:Sint}) should satisfy the variational equation $\delta\Lambda / \delta\Phi(\xi) = 0$, which cannot be solved analytically. Here we derive approximate solutions for some extreme values of the parameters $\sigma$ and $\gamma$. Our results in this subsection do not rely on the network ansatz, Eq.~(\ref{eq:network}), of the function $\Phi$.

\paragraph{Weak selection, $\gamma \to 0$}

In this limit, we can expand the integrand in Eq.~(\ref{eq:Sint}) to first order in $\gamma^2$, yielding
\begin{align}
\Lambda &\approx \sum_\mu p_\mu \log \bigg( 1 - \tfrac{\gamma^2}{2} \int \d\xi \, P(\xi|\varepsilon^\mu) \, \big( \Phi(\xi) - \psi^\mu \big)^2 \bigg) \nonumber \\
&\approx - \tfrac{\gamma^2}{2} \sum_\mu p_\mu \int \d\xi \, P(\xi|\varepsilon^\mu) \, \big( \Phi(\xi) - \psi^\mu \big)^2 \,,
\end{align}
where $P(\xi|\varepsilon^\mu)$ is the Gaussian distribution of $\xi$. To maximize the value of $\Lambda$, we set its variational derivative over the function $\Phi(\xi)$ to zero,
\begin{equation}
\frac{\delta\Lambda}{\delta\Phi(\xi)} = - \gamma^2 \sum_\mu p_\mu P(\xi|\varepsilon^\mu) \big( \Phi(\xi) - \psi^\mu \big) = 0 \,.
\end{equation}
Solving this equation yields
\begin{equation} \label{eq:weak}
\Phi^*(\xi) = \frac{\sum_\mu p_\mu P(\xi|\varepsilon^\mu) \psi^\mu}{\sum_\nu p_\nu P(\xi|\varepsilon^\nu)} = \frac{\sum_\mu p_\mu \psi^\mu \, \e^{- \frac{1}{2 \sigma^2} (\xi - \varepsilon^\mu)^2}}{\sum_\nu p_\nu \, \e^{- \frac{1}{2 \sigma^2} (\xi - \varepsilon^\nu)^2}} \,.
\end{equation}
This result can also be written succinctly as $\Phi^*(\xi) = \sum_\mu P(\varepsilon^\mu|\xi) \psi^\mu$, using Bayes' rule. The same expression has been derived in \cite{Xue2018}.

In the subcase where $\sigma$ is small, i.e., when the cue $\xi$ is accurate, the probability $P(\varepsilon^\mu|\xi)$ is nearly $1$ for the correct environment $\varepsilon^\mu$, hence the phenotypes are concentrated at the corresponding archetype $\psi^\mu$. This yields the tracking strategy. However, when $\sigma$ is large, i.e., when the cue is noisy, all environments $\varepsilon^\mu$ are likely; Eq.~(\ref{eq:weak}) becomes $\Phi^*(\xi) \approx \sum_\mu p_\mu \psi^\mu \equiv \bar{\psi}$, which means that an average phenotype $\bar{\psi}$ is produced regardless of the cue. This corresponds to the unvarying strategy.

\paragraph{Low noise, $\sigma \to 0$}

In this limit, the Gaussian distribution of $\xi$ in Eq.~(\ref{eq:Sint}) is concentrated near its mean, $\varepsilon^\mu$, so we can expand the integrand around that point. This yields, to first order in $\sigma^2$,
\begin{align}
\Lambda &\approx \sum_\mu p_\mu \log \bigg( \e^{- \frac{\gamma^2}{2} (\Phi(\varepsilon^\mu) - \psi^\mu)^2} \nonumber \\
&\quad \times \Big[ 1 - \tfrac{\sigma^2}{2} \Big( \gamma^2 \partial_a \Phi_i(\varepsilon^\mu) \, \partial_a \Phi_i(\varepsilon^\mu) + \cdots \Big) \Big] \bigg) \nonumber \\
&\approx - \tfrac{\gamma^2}{2} \sum_\mu p_\mu \Big[ \big( \Phi(\varepsilon^\mu) - \psi^\mu \big)^2 \nonumber \\
&\quad\; + \sigma^2 \Big( \partial_a \Phi_i(\varepsilon^\mu) \, \partial_a \Phi_i(\varepsilon^\mu) + \cdots \Big) \Big] \,.
\end{align}
This expression depends on the local values of the function $\Phi$ and its derivatives, $\Phi(\varepsilon^\mu)$, $\partial \Phi(\varepsilon^\mu)$, etc. To maximize $\Lambda$, we should have $\Phi^*(\varepsilon^\mu) \approx \psi^\mu$ and $\partial \Phi^*(\varepsilon^\mu) \approx 0$. It means that the ideal function $\Phi^*$ maps each environment $\varepsilon^\mu$ to its archetype $\psi^\mu$, and the mapping is locally ``flat'' --- the function value changes little in the neighborhood of $\varepsilon^\mu$. Since, for low noise, the cues $\xi$ are close to the actual environment $\varepsilon^\mu$, they will all be mapped to near the correct archetype $\psi^\mu$. This leads to the tracking strategy for any value of the selection strength $\gamma$.

\paragraph{High noise, $\sigma \to \infty$}

In this limit, the cue $\xi$ has a broad distribution that varies little with the environment $\varepsilon^\mu$, hence $P(\xi|\varepsilon^\mu) \approx P(\xi)$. As a result, the phenotype distribution will also be independent of the environment, and can be defined as
\begin{equation} \label{eq:pi}
\pi(\phi) \equiv \int d\xi \, P(\xi) \, \delta(\phi - \Phi(\xi)) \,.
\end{equation}
Using this phenotype distribution, the long-term growth rate $\Lambda$ can be written as
\begin{equation} \label{eq:Lambda-pi}
\Lambda \approx \sum_\mu p_\mu \log \int d\phi \, \pi(\phi) \, \e^{-\frac{\gamma^2}{2} (\phi - \psi^\mu)^2} \,.
\end{equation}
The distribution $\pi^*(\phi)$ that maximizes $\Lambda$ will constrain the ideal function $\Phi^*$ through Eq.~(\ref{eq:pi}).

Let us treat the subcases of small and large $\gamma$ separately. For a small $\gamma$, i.e., weak selection, we once again expand $\Lambda$ to first order in $\gamma^2$, which yields
\begin{align}
\Lambda &\approx \sum_\mu p_\mu \log \Big( 1 - \tfrac{\gamma^2}{2} \int d\phi \, \pi(\phi) \, (\phi - \psi^\mu)^2 \Big) \nonumber \\
&\approx - \tfrac{\gamma^2}{2} \sum_\mu p_\mu \int d\phi \, \pi(\phi) \, (\phi - \psi^\mu)^2 \nonumber \\
&= - \tfrac{\gamma^2}{2} \Big( \int d\phi \, \pi(\phi) \, \big( \phi - \bar{\psi} \big)^2 + \mathbb{V} [\psi] \Big) \,,
\end{align}
where $\mathbb{V}[\psi] = \sum_\mu p_\mu (\psi^\mu)^2 - \bar{\psi}^2$. From this expression it is clear that the optimal phenotype distribution is $\pi^*(\phi) = \delta(\phi - \bar{\psi})$, which agrees with the unvarying strategy found above.

For a large $\gamma$, it can be seen from Eq.~(\ref{eq:Lambda-pi}) that the distribution $\pi(\phi)$ should become sharply peaked at points where $\phi = \psi^\mu$. We can use the ansatz $\pi(\phi) = \sum_\mu \pi_\mu \, \delta(\phi - \psi^\mu)$, which is a discrete distribution with weights only at the archetypes $\psi^\mu$. Inserting this ansatz into $\Lambda$ yields
\begin{equation}
\Lambda \approx \sum_\mu p_\mu \log \pi_\mu \,.
\end{equation}
This expression recovers the model of bet-hedging (see, e.g., \cite{Rivoire2011}). The optimal values of $\pi_\mu$ are given by $\pi^*_\mu = p_\mu$. Therefore, the phenotype distribution will consist of separate peaks at each $\psi^\mu$, their relative sizes being proportional to the probability $p_\mu$ that each environment $\varepsilon^\mu$ occurs. To generate such a phenotype distribution, the function $\Phi^*(\xi)$ has to partition the environment space such that each partition has a total probability $p_\mu$.

\paragraph{Strong selection, $\gamma \to \infty$}

In this limit, the archetypes are far from one another as measured by the characteristic scale $1/\gamma$. Since a phenotype can be close to only one of the archetypes, there is a trade-off between the fitness values in different environments. In this case, the shape of the phenotype distribution can be understood by analyzing the geometry of the ``fitness set'' \cite{Levins1968, Mayer2017}.

Specifically, for each phenotype $\phi$, the fitness values $f_\mu(\phi) \equiv f(\phi;\varepsilon^\mu)$ for $\mu = 1, \cdots, m$ can be represented by a point in an $m$-dimensional fitness space. The collection of such points for all phenotypes $\phi$ forms the fitness set. Then, the average fitness of a population with a given phenotypic response function $\Phi(\xi)$ can be written as
\begin{equation} \label{eq:extended-fitness}
f_\mu[\Phi] \equiv \int d\phi \, \pi(\phi|\varepsilon^\mu) f_\mu(\phi) \,,
\end{equation}
where the phenotype distribution $\pi(\phi|\varepsilon^\mu)$ is given by
\begin{equation} \label{eq:pi|mu}
\pi(\phi|\varepsilon^\mu) \equiv \int d\xi \, P(\xi|\varepsilon^\mu) \, \delta(\phi - \Phi(\xi)) \,.
\end{equation}
The collection of those points, $\{ f_\mu[\Phi] \}$ for all possible phenotypic responses $\Phi(\xi)$, forms the ``extended fitness set''. Geometrically, each $f_\mu$ in the extended set can be considered as a linear combination of points from the original fitness set, weighted by the phenotype distribution in Eq.~(\ref{eq:extended-fitness}). By locating the point within the extended fitness set that maximizes the long-term growth rate, $\Lambda = \sum_\mu p_\mu \log f_\mu$, one can find the optimal phenotypic response and the phenotype distribution \cite{Mayer2017}.

\begin{figure}
\centering
\includegraphics[]{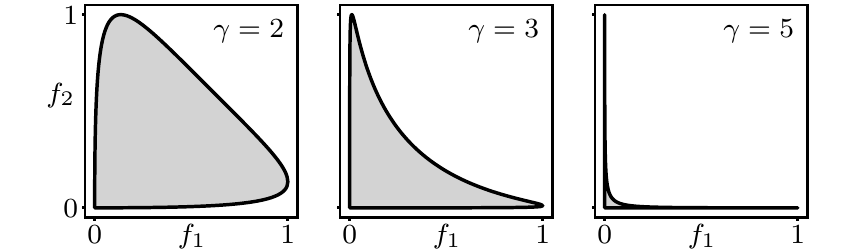}
\caption{\small Fitness sets (shaded area) for different values of the selection strength $\gamma$. Here $f_1$ and $f_2$ are fitness values of a phenotype in each of the two environments, with the corresponding archetypes separated by a distance $d = 1$.}
\label{fig:set}
\end{figure}

As an example, consider two environments, $\mu = 1, 2$. The fitness values are given by $f_1 = \e^{-\gamma^2 (\phi - \psi^1)^2 / 2}$ and $f_2 = \e^{-\gamma^2 (\phi - \psi^2)^2 / 2}$, where the two archetypes are assumed to be at a distance $d = 1$ without loss of generality. In this case, the fitness set is shown in Fig.~\ref{fig:set}. It can be seen that, when $\gamma \gg 1$, the fitness set is highly concave. As a result, the extended fitness set will be largely formed by linear combinations of points near the corners at $(1,0)$ and $(0,1)$. This means that the phenotype distribution mainly consists of phenotypes near the archetypes $\psi^1$ and $\psi^2$. Hence, regardless of the cue, the optimal phenotype distribution will be peaked at the archetypes.

%

\clearpage
\includepdf[pages={{},1,{},2,{},3,{},4}]{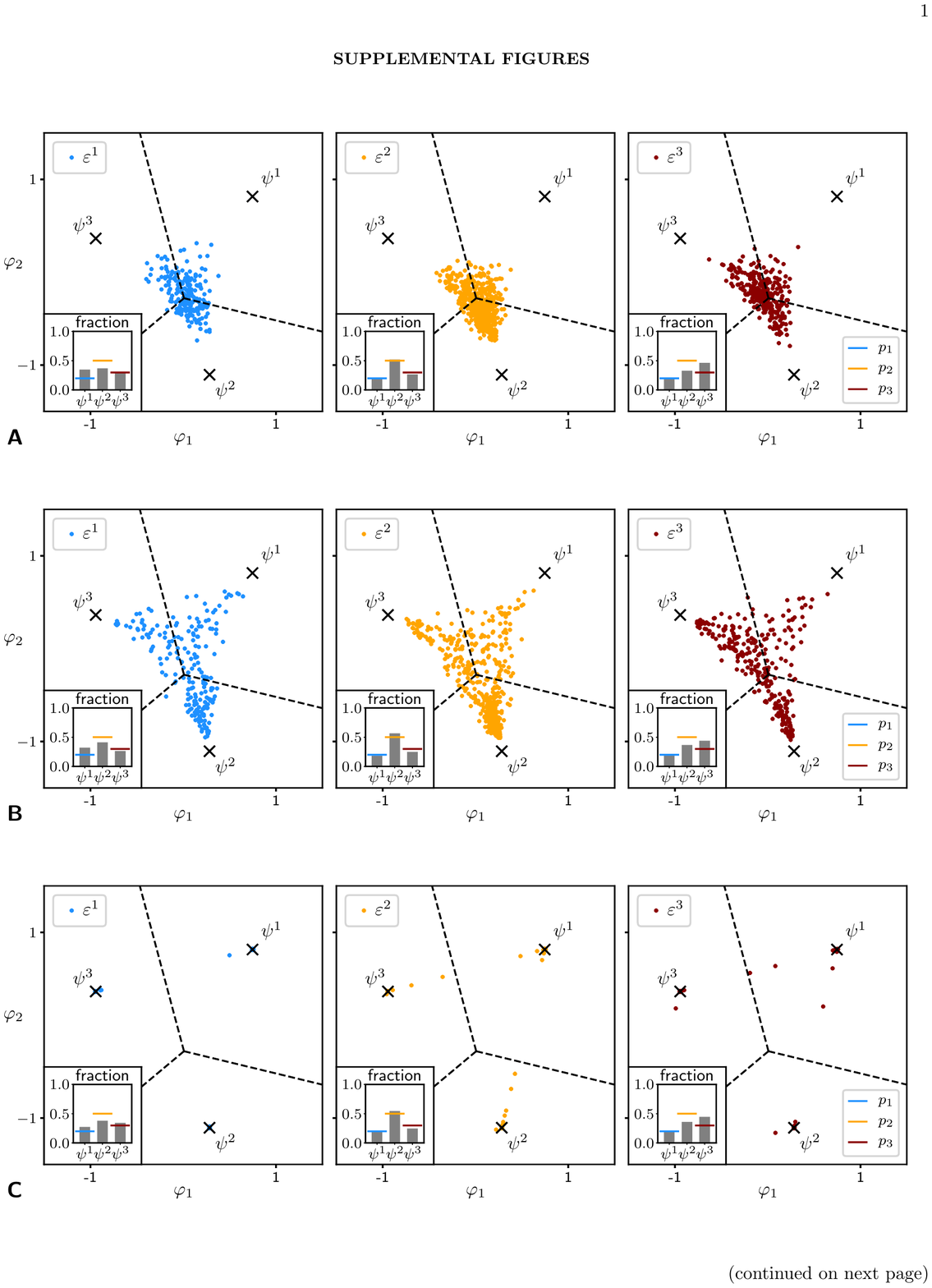}

\end{document}